\documentclass[twocolumn,showpacs,amsmath,amssymb,floatfix,prb]{revtex4}

\usepackage{graphicx}
\begin{document}

\title{Carbon Nanotube Thermal Transport: Ballistic to Diffusive}

\author{Jian Wang}
\author{Jian-Sheng Wang}

\affiliation{Department of Computational Science, National University
of Singapore, Singapore 117543, Republic of Singapore}

\date{21 October 2005}

\begin{abstract}
We propose to use $l_0/(l_0+L)$ for the energy transmission
covering both ballistic and diffusive regimes,
where $l_0$ is mean free path and $L$ is system length.
This formula is applied to heat conduction in
carbon nanotubes (CNTs).
Calculations of thermal conduction
show: (1) Thermal conductance at room temperature is
proportional to the diameter of CNTs for single-walled CNTs (SWCNTs)
and to the square of diameter for multi-walled CNTs (MWCNTs). (2)
Interfaces play an important role in thermal conduction in CNTs due to
the symmetry of CNTs vibrational modes. (3) When the phonon mean free
path is comparable with the length $L$ of CNTs in ballistic-diffusive
regime, thermal conductivity $\kappa$ goes as $L^{\alpha}$.  The
effective exponent $\alpha$ is numerically found to decrease with
increasing temperature and is insensitive to the diameter of SWCNTs
for Umklapp scattering process. For short SWCNTs ($<0.1\,\mu$m) we
find $\alpha \approx 0.8 $ at room temperature.  These results are
consistent with recent experimental findings.

\end{abstract}


\maketitle

Carbon nanotubes (CNTs) have been reported to have remarkable
electrical, mechanical, and thermal properties, making them the ideal
material for various engineering
applications.\cite{nanotubebook,nanotubereview} Of particular interest
in this letter is thermal conduction of CNTs. For example, the
anomalous enhancement of thermal conductivity has been experimentally
observed in CNTs composites\cite{enhance}.  However, many
conflicting results of CNTs thermal conduction have been reported
both theoretically and experimentally. The values of CNTs thermal
conductivity ranging from $ \rm 30\, W/mK $ to $ \rm 6600\, W/mK $
have been reported at room
temperature.\cite{thermalthousand,thermalhundred,thermaltens,Maruyamabao,yzhen,negative}
These drastically different values make it necessary to clarify which
is reliable and what are the conditions for obtaining such values.
Thermal conduction in CNTs may differ from the predictions of
Fourier's law based on bulk materials because the phonon mean free
path (MFP) can be comparable to the length of CNTs. Ballistic phonon
thermal transport behavior in CNTs has been observed in molecular
dynamic (MD) simulations\cite{Maruyamabao,yzhen} and recently in
experiments.  \cite{hychiu,ebrown} However, so far as we know, no
satisfactory theory has been able to cover both ballistic and
diffusive region.  Landauer formula \cite{gcrego} takes care of
ballistic thermal transport, while Boltzmann equations
\cite{gchen,mingo} cannot go over to purely ballistic transport.

In this letter, we propose a formula for thermal conduction covering
the range from ballistic to diffusive transport.  With this formula,
we discuss the CNTs thermal
conduction's dependence on the tube diameter and tube length, as well as
interface effect.

For thermal transport in quasi-one-dimensional system, the thermal
conductivity can be written as \cite{jwang}
\begin{eqnarray}
\label{12dconductance}
 \!\!\! \kappa \!\! &=& \!\!\frac{L}{S} \!\sum \limits_{n,v_n>0} \int \frac{d{q}}{2\pi }\hbar
 \omega_{n}({q})v_{n}({q}) \! \frac{\partial f}{\partial T} \mathcal{T}_{n}({q},
 \omega_{n}),
\end{eqnarray}
where $ L$ and $S $ are the length and cross-section area of the
system, $\mathcal{T}_{n}({q}, \omega_{n})$ the energy transmission for
the $n$-th branch wave at longitudinal momentum ${q}$, angular
frequency $\omega_{n}$, and $f(\omega_n, T)$ is the Bose-Einstein
distribution at temperature $T$.  $v_{n}$ is group velocity along the
direction of thermal transport given by $v_{n}=\partial \omega_n
/\partial q $.  The integration over momentum ${q}$ is within 
the first Brillouin zone.  The central problem is to calculate the
energy transmission $\mathcal{T}_{n}({q}, \omega_{n})$. We propose
that energy transmission can be calculated for three different thermal
transport regimes according to the relation between the MFP
$l_0(q)$ and the length of the system $L$. (1) \textsl{ Ballistic
regime } $l_0 \gg L$. The MFP is much larger than the length of the
system. Thermal transport in this regime can be described by Landauer
quantum transport formula. \cite{jwang} There are two equivalent
methods for calculating energy transmission resulting from boundary,
conjunction, or defect scattering from atomistic point of view: the
scattering boundary mode match method \cite{jwang} and the
non-equilibrium Green function method. \cite{mingo2} (2)\textsl {
Ballistic-diffusive regime} where $l_0 \sim L$.  The MFP is comparable
with the length of the system. In this regime, the energy transmission
is approximated by $\mathcal{T}_{n}({q}, \omega_{n}) =
{l_0}/{(L+l_0)}$, as has been shown in Ref.~\onlinecite{qtranport} for
electronic transport. (3) \textsl{ Diffusive regime} $l_0 \ll L$. The
MFP is much smaller than the length of the system. In this regime,
$\mathcal{T}_{n}({q}, \omega_{n}) \simeq {l_0}/{L}$. So thermal
conductivity in this limit is given by the well-known
Boltzmann-Peierls formula $ \kappa = \frac{1}{S} \sum_n \limits \int
\frac{d{q}}{2\pi }\hbar \omega_{n}({q})v_{n} \frac{\partial
f}{\partial T} l_0^{n}(q)$.

\begin{figure}[bt]
\includegraphics[width=1.00\columnwidth]{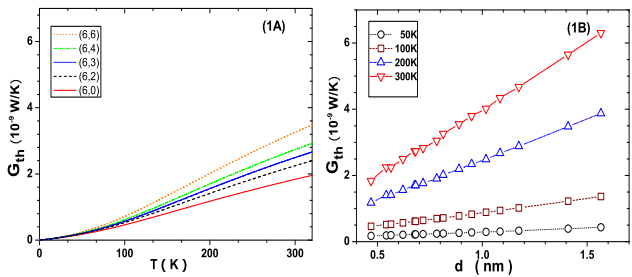}
\caption{\label{fig:thermalconductance} Thermal Conductance for
SWCNTs.  \textbf{(1A)} Different chirality SWCNTs as a function of
temperature.  \textbf{(1B)} Different diameter SWCNTs at
$T=50,\,100,\,200,\,300\,$K respectively. }
\end{figure}

\emph{Diameter dependence.} Here we assume that the thermal transport
in CNTs is in the ballistic regime.  To simplify the discussion and
catch the essential character of ballistic thermal transport in CNTs,
we assume the energy transmission $\mathcal{T}_{n}({q},
\omega_{n})\thickapprox 1$. This assumption is justified for short
CNTs with few imperfections at moderate
temperatures\cite{hychiu,yzhen}.  Several different definitions for the
cross section for CNTs have been used in
Ref.~\onlinecite{thermalthousand,thermalhundred,thermaltens,Maruyamabao,yzhen,
negative}, because it is not well defined.  To circumvent this
ambiguity, we would like to use more experimentally oriented physical
quantities: thermal conductance $G_{{\rm th}}=\kappa S/L$ to measure
the property of thermal conduction of CNTs.  In our computation,
phonon dispersion for single-walled CNTs (SWCNTs) is calculated using
the method and force constants in Ref.~\onlinecite{nanotubebook}.  The
group velocity is calculated from the phonon dispersion by the method
in Ref.~\onlinecite{jwang}. The results of thermal conduction for
different chirality and diameter SWCNTs at different temperatures are
illustrated in Fig.~\ref{fig:thermalconductance}. It can seen from
Fig.~\ref{fig:thermalconductance} that at low temperature, for example
$T=50\,$K, thermal conductance almost does not change with diameter or
chirality. At the extremely low temperature limit, all CNTs give the
quantized universal thermal conductance \cite{ebrown} $4 \pi^2
k_B^2T/3h$, which corresponds to the four channels of quantum
transport for acoustic phonons. When temperature increases, more
optical phonon channels will contribute to thermal transport. However,
it is contrary to the intuitive conclusion \cite{ebrown} that thermal
conductance at high temperature $T$ will be proportional to the
total numbers $N_{{\rm ph}}$ of phonon channels in CNTs. Thermal
conductance at high temperature is proportional to the diameter $d$
of CNTs, irrespective of their chirality.  For example, the ratio of
channel number SWCNTs $(6,4)$ to $(6,0)$ is $N^{(6,4)}_{{\rm
ph}}/N^{(6,0)}_{{\rm ph}}\approx 6.33$, while their diameter ratio is
$d^{(6,4)}/d^{(6,0)}\approx 1.45$.  It can be seen from
Fig.~\ref{fig:thermalconductance} that the ratio of thermal
conductance at $300\,$K is about $1.45$, not $6.33$.
This proportionality to diameter is implicitly demonstrated
in Ref.~\onlinecite{mingo} by counting the number of phonon branches
at frequency $\omega$ to calculating the transmission.  This simple
result can be easily understood if we approximate the integrand in
Eq.~(\ref{12dconductance}) by a constant, obtaining the result that
$G_{{\rm th}}$ is proportional to the total number of modes divided by
the unit cell length, which is proportional to $d$.  For MWCNTs, if
thermal conductance above certain temperature can be added from the
contribution of each shell independently, the total thermal
conductance will be $\varpropto \int r dr \varpropto d^2$. The
dissipation power results in Ref.~\onlinecite{hychiu} have a tendency
of $d^2$. It should be noted that the above results only characterize
the thermal conductance in pure ballistic regime.  In actual
experiments, the situation is more complicated for thermal conductance
in sub-micrometer lengths. For example, the contact interface and
phonon scattering are inevitable in experiment. We will discuss these
effects on thermal conduction in CNTs.

\emph{Interface effect.}  Interface effect on CNTs thermal transport
has been studied through effective medium theory \cite{cwnan} and
empirical random walk simulation. \cite{hmduong} However they are only
rough models, not from first principles.  Here we consider the
interface effect from a lattice point of view. We consider an extreme
case, where the semiconductor nanotube junction structure (11,0) and
(8,0) is first constructed by a geometrical method as in
Ref.~\onlinecite{nanotubebook}. The structure is optimized by a
second-generation Brenner potential \cite{dwbrenner} to let the atoms
get their equilibrium positions. The force constants are derived
numerically from the same potential.  The phonon dispersion is
calculated from these linearized force constants.  Four acoustic
branches are considered for energy transport: the longitudinal mode
(LA), doubly degenerate transverse mode (TA), and the unique twist
mode (TW) in nanotubes. Following the scattering boundary method,
\cite{jwang} we calculated the energy transmission across the CNTs
conjunction. The details of calculation are presented in
Ref.~\onlinecite{jwang}.  The reflected and the transmitted waves
across the junction for the incident LA mode waves are both only LA
modes. We think that this is due to the high symmetrical properties of
atomic motion for nanotubes.
The LA and TA modes are common
symmetrical motions for both the left and right lead. So they propagate
through the conjunction. In contrast, the transmissions of the TW mode
and many other optical modes are nearly zero or very small.  From
these results, we consider the interface between CNTs and substrate
will influence the experimental measured CNTs thermal conductivity
greatly, which has been indicated in experimental results
\cite{negative,hychiu}. We argue that chemical functional
reorganization of the CNTs ends or proper choice of the more
symmetry-shared substrate matrix will improve thermal transport of
CNTs.

\emph{Length effect.} Purely ballistic thermal transport will make
thermal conductivity $ \kappa $ diverge linearly with the length $L$
of CNTs. Anomalous thermal transport for 1D nonlinear lattice predicts
\cite{lepri} that $ \kappa $ will diverge with $L$ as $ \kappa
\varpropto L^{\alpha}$.  For 1D nonlinear lattice with transverse
motion, which is similar to CNTs, it is found $\alpha =1/3$ through MD
simulation and mode-coupling theory. \cite{Jswang} A few results of
tube length effect on thermal transport in CNTs have been reported
recently through MD simulation\cite{Maruyamabao,yzhen}.  A length
dependence of the thermal transport in CNTs is also observed in
experiments\cite{hychiu}. It can be seen from
Eq.(\ref{12dconductance}) that thermal conductivity will depend on the
length of CNTs in ballistic-diffusive regime.
\begin{figure}[b]
\includegraphics[width=1.00\columnwidth]{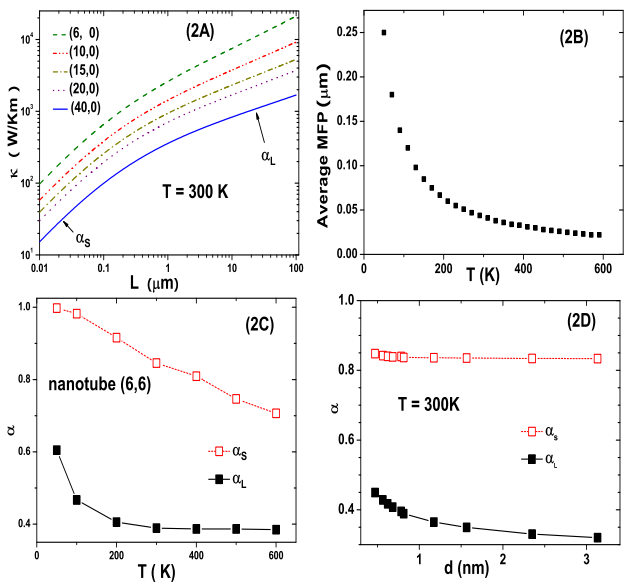}
\caption{\label{fig:mfpfigure} Length divergence in
ballistic-diffusive regimes: \textbf{(2A)} Thermal conductivity for
different chiralities SWCNTs at 300\,K; \textbf{(2B)} Average mean free
path length for nanotube (6,6) at different temperatures;
\textbf{(2C)} Change of power law exponent for short and long CNTs
with temperature; \textbf{(2D)}Change of power law exponent for short
and long CNTs with diameter;}
\end{figure}

The problem to calculate thermal conductivity from
Eq.~(\ref{12dconductance}) in ballistic-diffusive regime is to
compute MFP, which depends on frequency and wave length of the
vibration modes, as well as temperature. Assuming that the
scattering mechanisms are not coupled, MFP can be written
\cite{Klemens} as $ 1/l_0=\sum_{i}1/l_i$, where the index $i$
denotes various interaction phonon processes. In CNTs, the phonon
scattering process may be different from that of bulk and is not
well known. Here, for simplicity, we first consider the Umklapp
scattering. Umklapp phonon scattering $l_U^{-1}$ in CNTs has been
shown linear $T$ dependence at high temperatures through
experiment \cite{negative} and is usually given a $\omega^2$
dependency on the frequency. \cite{carbon} So we take the form $
l_{U}=A/(\omega^2T) $ for Umklapp scattering in CNTs, where $A$ is
taken to be the same as derived for graphene in
Ref.~\onlinecite{carbon}. The phonon dispersion is calculated as
in Ref.~\onlinecite{nanotubebook}.  The cross-section area $S$ is
defined as $S=\pi d^2$/4. Results are illustrated in
Fig.~\ref{fig:mfpfigure} through Eq.~(\ref{12dconductance}) for
ballistic-diffusive regime in CNTs. We should note that the
exponent $\alpha$ in Fig.~\ref{fig:mfpfigure}(C,D) is an effective
exponent in a limited length region, not the asymptotic one.
\cite{lepri,Jswang} It can be seen from Fig.~2A that thermal
conductivity shows different length dependence scaling
$L^{\alpha}$ for short CNTs $(<0.1\mu m )$ and long CNTs $(>10 \mu
m )$, denoted as $\alpha_S$ and $\alpha_L$ respectively, when
Umklapp scattering in CNTs is considered in ballistic-diffusive
regime. This is due to the relation between MFP and the length of
CNTs. Average MFP $\bar l_0(T) = \int l_0(\omega,T)D(\omega)d
\omega $ is calculated from the phonon dispersion relation for CNT
(6,6), where $D(\omega)$ is normalized density of states,
satisfying $\int D(\omega)d \omega =1$.
Figure~\ref{fig:mfpfigure}B indicates that average MFP decreases
rapidly from $0.25\,\mu$m to $0.06\,\mu$m, when temperature goes
up from $50\,$K to $200\,$K. We also calculated average MFP for
other chirality CNTs and found little difference from that of CNT
(6,6). Fig.~2C shows the change of $\alpha_S$ and $\alpha_L$ with
temperature. For example, when $T=50\,$K, the average MFP is as
large as $0.25\, \mu$m and so the $\alpha_S$ is almost equal to
$1$.  When temperature increases, the MFP becomes shorter, so
$\alpha_S$ and $\alpha_L$ both decrease with temperature. The
dependence of $\alpha_S$ and $\alpha_L$ on CNTs diameter is
illustrated in Fig.~2D. $\alpha_S$ almost does not change with the
chirality of CNTs and gives $\alpha_S \sim 0.83 $ at room
temperature, while $\alpha_L$ decreases slowly with increase of
the diameter of CNTs, and gives its value about $\alpha_L \sim
0.35 $ for CNTs with diameter larger than $1 n$m at room
temperature. This kind of temperature dependence and chirality
dependence agrees with MD simulation results.
\cite{Maruyamabao,yzhen} It is worth noting that, for CNTs with
length less than $0.1\, \mu$m, $\alpha = 0.32$ $ \hbox{and}$
$\alpha = 0.40 $ for CNT (5,5) at room temperature are reported by
non-equilibrium MD simulation\cite{Maruyamabao}, while $ \alpha
\approx 1$ is observed using equilibrium MD
simulation\cite{yzhen}.  For MWCNT with length less than $0.5\,
\mu$m, roughly $\alpha \sim 1 $ is reported. \cite{hychiu} We
think that the difference for non-equilibrium MD and equilibrium
MD comes from the boundary condition because in short CNTs the
result should be sensitive to the boundaries.

Next we discuss other phonon scattering mechanisms. If the frequency
dependence of MFP is assumed as $1/l_0 \varpropto \omega ^{r}$, then
at high temperature the length dependence of thermal conductivity
given by Eq.~(\ref{12dconductance}) can be estimated as $\kappa \sim
\int \frac{1}{1/l_0+1/L}d \omega \varpropto L^{\frac{r-1}{r}} $, that
is $\alpha = {(r-1)}/{r}$.  For Umklapp scattering process, $r=2$ and
$\alpha = 0.5$, which corresponds to the effective exponent at length
about $1 \mu$m.  For defect scattering mechanism, we have $1/l_0
\varpropto \omega ^{4}$, $\alpha = 0.75$.  Thus, it can be argued that
strong frequency dependence scattering (large $r$) will not have much help
in eliminating thermal conductivity divergence, while weak frequency
dependence scattering will contribute to eliminating this
divergence. Finally, in the very long tube limit, $\kappa$ approaches
a constant within our theory.

\emph{Conclusions.}
Crossover in thermal conduction from ballistic to diffusive is
illustrated.
Thermal conductance at room temperature is found $\varpropto d$
for SWCNTs and $\varpropto d^2$ for MWCNTs.  Interfaces plays an
important role in thermal conduction in CNTs due to the high
symmetrical property of CNTs vibration modes. Possible ways to
improve thermal conduction in CNTs are suggested.  In
ballistic-diffusive regime, thermal conductivity $\kappa$
behaves as $L^{\alpha}$. The effective exponent $\alpha$ is
numerically found decreasing with
increasing temperature and insensitive to the diameter of SWCNTs
for Umklapp scattering process. The possible mechanism for the
reduction of divergence for thermal conductivity is also
discussed.  Although the formula  is still
phenomenological, it does cover ballistic and diffusive regimes with a
smooth crossover and gives a reasonably simple picture in the whole
temperature range.
This work is supported in part by a Faculty Research
Grant of National University of Singapore.

\end{document}